\documentclass[twocolumn,aps,prl,showpacs]{revtex4}
\usepackage{graphicx}
\DeclareGraphicsExtensions{ps,eps,epsf}
\newcommand{ \ybco }{\mbox{YBa$_2$Cu$_3$O$_{6+x}$}}
\newcommand{ \ybcoo }{\mbox{YBa$_2$Cu$_3$O$_{6.92}$}}
\newcommand{ \yezv }{\mbox{YBa$_2$Cu$_4$O$_{8}$}}
\newcommand{ \ycado }{\mbox{Y$_{0.8}$Ca$_{0.2}$Ba$_2$Cu$_3$O$_{6.95}$}}
\newcommand{ \lasco }{\mbox{La$_{2-x}$Sr$_x$CuO$_{4}$}}

\begin{document}

\title{Oxygen superstructures throughout the phase diagram of $\bf (Y,Ca)Ba_2 Cu_3 O_{6+x}$}
\author{J. Strempfer$^1$, I. Zegkinoglou$^1$, U. R\"utt$^1$, M. v. Zimmermann$^2$,
C. Bernhard$^1$, C. T. Lin$^1$, Th. Wolf$^3$,
and B. Keimer$^1$}
\affiliation{$^1$ Max-Planck-Institut f\"ur Festk\"orperforschung, Heisenbergstr. 1,
D-70569 Stuttgart, Germany}
\affiliation{$^2$ Hamburger Synchrotronstrahlungslabor HASYLAB at
Deutsches Elektronen-Synchrotron DESY, Notkestr. 85, D-22603 Hamburg, Germany}
\affiliation{$^3$ Forschungszentrum Karlsruhe, IFP, D-76021 Karlsruhe, Germany}

\date{\today}

\begin{abstract}
Short-range lattice superstructures have been studied with
high-energy x-ray diffuse scattering in underdoped, optimally
doped, and overdoped $\rm (Y,Ca)Ba_2 Cu_3 O_{6+x}$. A new
four-unit-cell superstructure was observed in compounds with
$x\sim 0.95$. Its temperature, doping, and material dependence was
used to attribute its origin to short-range oxygen vacancy
ordering, rather than electronic instabilities in the $\rm CuO_2$
layers. No significant diffuse scattering is observed in
\mbox{YBa$_2$Cu$_4$O$_{8}$}. The oxygen superstructures must be
taken into account when interpreting spectral anomalies in
$\rm (Y,Ca)Ba_2 Cu_3 O_{6+x}$.
\end{abstract}


\maketitle

 The two-dimensional strongly correlated electron system
in the layered copper oxides is known to be susceptible to at
least two types of instabilities at nonzero doping: high
temperature superconductivity, and ``stripe'' ordering of spin and
charge degrees of freedom \cite{Kiv02}. The question of whether
fluctuating stripes are a prerequisite for high temperature
superconductivity remains one of the central unanswered questions
in the field. Several experimental techniques are suitable as
probes of stripe order and fluctuations, including x-ray
scattering, magnetic and nuclear neutron scattering, NMR and NQR,
and scanning tunneling spectroscopy \cite{Kiv02}. X-rays couple
directly to the charge, and the high photon energy ensures that
both static charge ordering and charge excitations up to high
energies can be detected. Notably, x-ray superstructure
reflections due to static stripe ordering were observed in
Nd-substituted \lasco\ \cite{vZi98} following an initial
observation by neutron diffraction \cite{Tra95a}. Furthermore, a
recent x-ray scattering study of underdoped \ybco\ has uncovered
diffuse features whose temperature dependence was reported to
exhibit an anomaly around the ``pseudogap'' temperature
\cite{Isl02}. This anomaly was interpreted as a signature of
electronic stripe formation. In this system, however,
superstructures due to oxygen ordering with wave vectors depending
sensitively on the oxygen content are also observed \cite{And99}.
As both phenomena are associated with lattice distortions and are
thus expected to be intimately coupled, it is difficult to
establish which features of the x-ray data originate in short
range oxygen ordering, and which can be attributed to electronic
stripe ordering or fluctuations.

In order to answer this question unambiguously, we have
investigated the x-ray diffuse intensity in underdoped, optimally
doped, and overdoped $\rm (Y,Ca)Ba_2 Cu_3 O_{6+x}$ single
crystals. Surprisingly, diffuse features with a well defined
four-unit-cell periodicity were observed even in optimally doped
\mbox{YBa$_2$Cu$_3$O$_{6.92}$} where the density of oxygen
vacancies is low. However, the diffuse features were observed to
depend only on the oxygen concentration x in \ybco\, and not on
the charge carrier concentration in the CuO$_2$ planes which was
varied independently from the oxygen content through Ca
substitution. In addition, no significant diffuse scattering was
observed in {\mbox{YBa$_2$Cu$_4$O$_{8}$}, a naturally underdoped
material that does not sustain oxygen defects. The diffuse
intensity therefore arises from short-range oxygen ordering and
associated lattice distortions, and any signatures of stripe
ordering or fluctuations must be much weaker. The oxygen
superstructure induces substantial lattice deformations in the
CuO$_2$ layers and must be taken into account when interpreting
phonon anomalies in this material \cite{Mook98,Pin03}.

The experiments were conducted at the high-energy wiggler
beamlines BW5 at the Hamburger Synchrotronstrahlungslabor at the
Deutsches Elektronen-Synchrotron and 11-ID-C at the Advanced
Photon Source at the Argonne National Laboratory. The x-ray
energies were 100 keV and 115 keV, respectively. Both beamlines
were optimized for providing as much flux as possible for the
broad diffuse reflections. This was achieved by using a broad
energy bandwidth. At 11-ID-C, a Si-$(3, 1, 1)$ monochromator is
located almost at the 1:1 position  between beam source (wiggler)
and sample position. Therefore, the entire 10 mm width of the Laue
monochromator crystal can be used for focusing on the sample
\cite{Rut01}. At BW5, a Si-Ge-gradient crystal diffracting a large
energy bandwidth due to the lattice parameter variation was used
as monochromator \cite{Kei98}.

\begin{figure}
\includegraphics[width=0.95\linewidth]{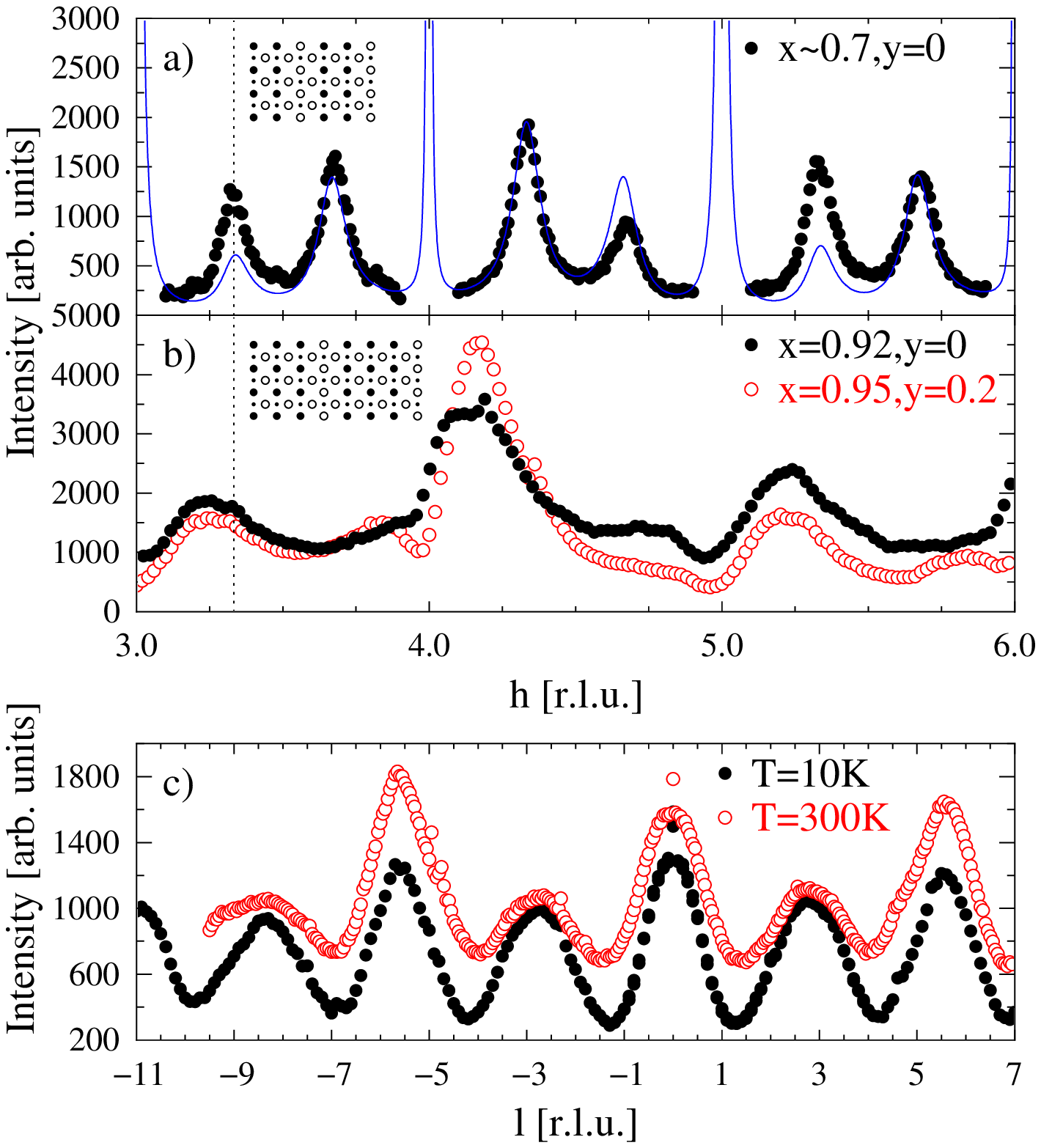}
\caption{\label{fig:hscans} Diffuse scattering intensities in
$\mbox{Y$_{1-y}$Ca$_{y}$Ba$_2$Cu$_3$O$_{6+x}$}$. (a) Scans along
$(h,0,2.5)$ for the ortho-III phase of
{\mbox{YBa$_2$Cu$_3$O$_{6.75}$}}, together with a simulation
(solid line) based on the structure found by Plakhty et al.
\cite{Pla95}. (b) $(h,0,5.6)$ scan for both
{\mbox{YBa$_2$Cu$_3$O$_{6.92}$}} ($\bullet$) and
{\mbox{Y$_{0.8}$Ca$_{0.2}$Ba$_2$Cu$_3$O$_{6.95}$}} ($\circ $) (c)
$l$-scans of {\mbox{YBa$_2$Cu$_3$O$_{6.92}$}} at both T=10K and
T=300K along $(4.25,0,l)$. The insets in a) and b) show the
ortho-III and ortho-IV patterns, respectively, where large full
(open) circles denote occupied (unoccupied) oxygen-sites.}
\end{figure}

The samples studied included one optimally doped \ybco\ crystal
with $x\sim 0.92$ and $\rm T_c = 92.7$ K, two underdoped \ybco\
crystals with $x\sim 0.75$ and $\rm T_c =67$ K and $x\sim 0.65$
and $\rm T_c =60$ K,  respectively, and a highly overdoped crystal
of composition {\mbox{Y$_{0.8}$Ca$_{0.2}$Ba$_2$Cu$_3$O$_{6.95}$}
and $\rm T_c = 73$ K \cite{Lin02}. The crystal volumes were
approximately $2\times 2 \times 0.4$ mm$^3$, and the optimally and
overdoped crystals were fully detwinned. In addition, one
untwinned {\mbox{YBa$_2$Cu$_4$O$_{8}$} crystal of approximate
volume $0.5 \times 0.8 \times 0.1$ mm$^3$ and $\rm T_c =81$ K was
investigated. In the following, the wave vector components
$(h,k,l)$ are indexed in the orthorhombic space group $Pmmm$ for
\ybco\ and $Ammm$ for \yezv. The lattice parameters are
a=3.8158(1)\AA, b=3.8822(1)\AA, and c=11.6737(3)\AA\ for
{\mbox{YBa$_2$Cu$_3$O$_{6.92}$} at $T=270$ K and a=3.8410(3)\AA,
b=3.8720(3)\AA\ and c=27.231(2)\AA\ for \yezv. The samples were
mounted in closed-cycle cryostats capable of reaching temperatures
between 10 and 300 K or between 20 and 500 K, depending on the
experiment. The cryostats were mounted on four-circle goniometers
to allow a wide coverage of reciprocal space. Due to the high
photon energy, the x-ray penetration depth is comparable to the
sample dimensions. Strain effects in the near-surface region,
which can influence x-ray scattering measurements with lower
photon energies, are thus not relevant here.

Fig. \ref{fig:hscans} shows the intensities of x-rays scattered
from the underdoped, optimally doped and overdoped
$\rm (Y,Ca)Ba_2 Cu_3 O_{6+x}$ single crystals described above. The diffuse
features shown are $\sim 5$ orders of magnitude weaker than the
main Bragg reflections. The intensity is peaked at wave vectors
$h= 0.33$ for \mbox{YBa$_2$Cu$_3$O$_{6.75}$}, and $h=0.25$ for
\mbox{YBa$_2$Cu$_3$O$_{6.92}$} and
{\mbox{Y$_{0.8}$Ca$_{0.2}$Ba$_2$Cu$_3$O$_{6.95}$} (Figs.
\ref{fig:hscans}a-b), indicating short-range superstructures with
periodicities equal to three and four elementary orthorhombic unit
cells, respectively. As a function of the wave vector component
$l$ perpendicular to the copper oxide layers (Fig.
\ref{fig:hscans}c), the intensity shows a modulation with a
periodicity characteristic of interatomic distances within the
unit cell, in particular the distances between the copper oxide
chain and plane layers and between the chain layer and the apical
oxygen, as described in \cite{Isl02}. The diffuse scattering thus
results from lattice displacements encompassing the entire unit
cell.

While the four-cell superstructure in optimally and overdoped
$\rm (Y,Ca)Ba_2 Cu_3 O_{6+x}$ has thus far not been reported, the
three-cell superstructure in underdoped $\rm YBa_2 Cu_3 O_{6.75}$
is known as the ``ortho-III phase'' \cite{And99}. Its primary
origin is an ordering of oxygen defects in a pattern in which one
empty CuO chain follows two fully occupied chains (inset in Fig.
\ref{fig:hscans}a). Due to the oxygen diffusion kinetics, the
ortho-III phase is always short-range ordered, and the width of
the superlattice reflection of $\Delta h = 0.12(1)$ r.l.u. we
observe compares favorably with prior work \cite{And99,Pla95}. The
measured intensities are also in good agreement with the
literature, although some deviations are found (Fig.
\ref{fig:hscans}a). In particular, the intensity difference of the
satellites with wave vectors $\pm 0.33$ surrounding the main Bragg
reflections is well reproduced by a calculation based on the
atomic displacements given in Ref. \cite{Pla95}. Similar
observations were made on the ``ortho-V phase'' of the
$\rm YBa_2 Cu_3 O_{6.65}$ crystal (not shown).

We now turn to the newly discovered four-cell superstructures in
optimally doped $\rm YBa_2 Cu_3 O_{6.92}$ and highly overdoped
{\mbox Y$_{0.8}$Ca$_{0.2}$Ba$_2$Cu$_3$O$_{6.95}$}. The
periodicities, intensities and correlation lengths of the
superstructures in both samples are virtually identical within the
statistical accuracy of the data. (The range around the $(4,0,0)$
position in Fig. \ref{fig:hscans}b has to be disregarded, because
a tail of the main Bragg reflection obstructs the diffuse
intensity.) Charge density wave or stripe correlations are
expected to be characterized by a wave vector that depends
strongly on the charge carrier concentration, and by an amplitude
that is strongly reduced in heavily overdoped samples. As the
charge carrier concentration in the $\rm CuO_2$ layers is very
different in the two samples (whereas their oxygen content is
nearly identical), we conclude that charge order or fluctuations
within the layers can at most give a minor contribution to the
diffuse intensity.

\begin{figure}
\includegraphics[width=0.95\linewidth]{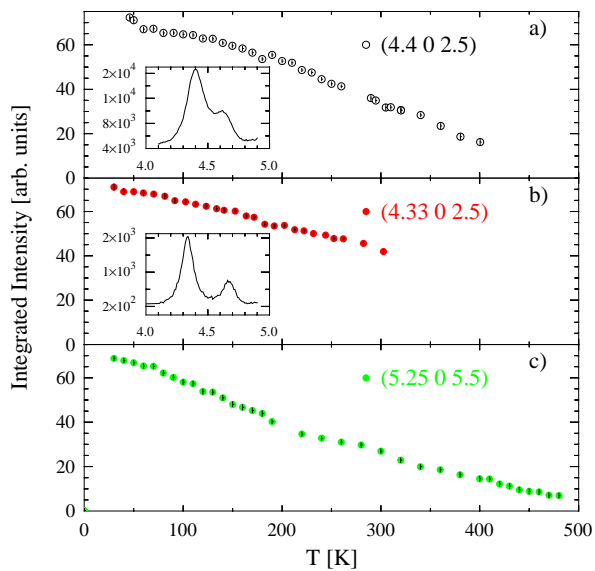}
\caption{\label{fig:tabh} Temperature dependences of the
integrated intensity of (a) the $(4.4,0,2.5)$/$(4.6,0,2.5)$
reflections in $\rm YBa_2 Cu_3 O_{6.65}$, (b) the
$(4.33,0,2.5)$/$(4.66,0,2.5)$ reflections in $\rm YBa_2 Cu_3
O_{6.75}$ and the $(5.25,0,5.6)$ reflections in \ybcoo. The insets
in (a) and (b) show the respective Q-scan at low temperature.}
\end{figure}

In analogy to the underdoped samples discussed above, we hence
attribute the diffuse features in $\rm (Y,Ca)Ba_2 Cu_3 O_{6+x}$
samples with $x \sim 0.95$ to an ``ortho-IV'' oxygen-ordered phase
characterized, on average, by a sequence of three full and one
empty CuO chain (inset in Fig. \ref{fig:hscans}b). Because of the
low density of oxygen defects, islands of the ortho-IV phase are
expected to be small and dilute. This explains the large width of
the superstructure peaks of $\Delta h = 0.29(1)$ r.l.u. in this
compound. A complementary way of looking at these data is through
an analysis in terms of pair correlations between oxygen defects,
which leads to a Fourier series for the diffuse intensity
\cite{Cow50}. In fact, a single Fourier component corresponding to
an enhanced probability for occupation of oxygen defects on
third-nearest-neighbor chains already provides a good description
of the width of the diffuse features in the $(h,0,0)$ direction:
$I(h)\sim 1-\alpha\cos(4 \pi h)$. This confirms the short range
nature of the oxygen ordering. A quantitative description of the
correlation coefficient $\alpha$ in different Brillouin zones will
require a more comprehensive simulation of the lattice
distortions, which is beyond the scope of this paper. The salient
features of the diffuse intensity, including especially the
intensity asymmetry of the satellite features above and below the
main Bragg reflections (Fig. \ref{fig:hscans}b), are, however,
similar to those observed in the ortho-III ordered sample, which
implies similar atomic displacements.

The temperature dependences of the integrated intensities of the
ortho-V, ortho-III and ortho-IV peaks are shown in Figs.
\ref{fig:tabh}a-c. Whereas for the ortho-V and ortho-III samples
scans along $h$ were performed, as shown in the insets, the
integrated intensity of the optimally doped sample was obtained
from $l$-scans over the diffuse peaks, as shown in Fig.
\ref{fig:hscans}c. This allows the most reliable subtraction of
the temperature dependent background from thermal diffuse
scattering. These temperature dependences further confirm that all
features originate in oxygen ordering rather than charge
instabilities in the $\rm CuO_2$ layers. Independent of the doping
level, the superstructures persist well above room temperature,
with no anomalies observed at the superconducting transition
temperature or other temperatures associated with the onset of
electronic instabilities. In particular, our data are at variance
with Islam {\it et al.}'s claim of an anomalous temperature
dependence of the ortho-V reflections near the onset of the
electronic ``pseudogap'' \cite{Isl02}. The intensity variation
observed in Ref. \cite{Isl02} is clearly outside the statistical
variation of the data of Fig. \ref{fig:tabh}a. Rather, the
intensity of the diffuse features is reduced smoothly upon heating
and vanishes around 400-450 K. Prior work in samples with $0.5
\leq x \leq 0.8$ has shown that in this temperature range, oxygen
order is obliterated due to progressively rapid oxygen diffusion
\cite{And99}.

\begin{figure}
\includegraphics[width=0.95\linewidth]{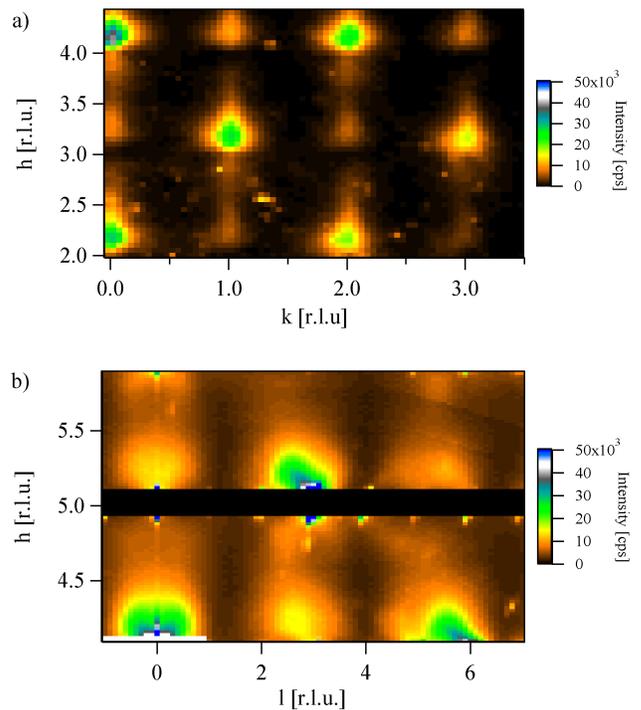}
\caption{\label{fig:ycado} Contour plots of the diffuse intensity
in the (a) $(h,k,5.5)$-plane and (b) $(h,0,l)$-plane of \ycado.
The main Bragg reflections are masked.}
\end{figure}

\begin{figure}
\includegraphics[width=0.95\linewidth]{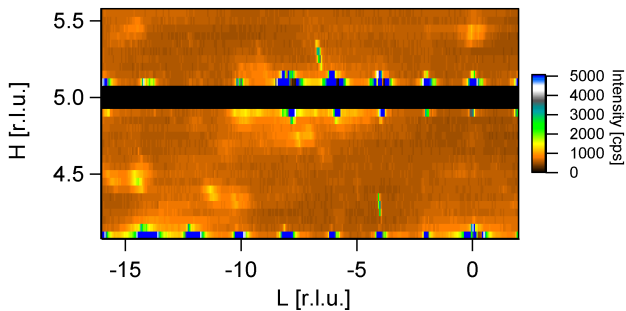}
\caption{\label{fig:y124} Contour plot of the diffuse intensity in
the $(h,0,l)$ scattering plane of \yezv.}
\end{figure}

Fig. \ref{fig:ycado} shows maps of the diffuse intensity in the
$(h, k,5.5)$ and $(h,0,l)$ scattering planes for the {\mbox
Y$_{0.8}$Ca$_{0.2}$Ba$_2$Cu$_3$O$_{6.95}$} sample. Similar maps
for the {\mbox YBa$_2$Cu$_3$O$_{6.92}$} sample (not shown) are
nearly indistinguishable from those of Fig. \ref{fig:ycado}, in accord with
Fig. \ref{fig:hscans}b. Panel a shows that the superstructure is
only observed along the a-direction, that is, perpendicular to the
CuO chains. Diffuse features induced by charge density wave
ordering along the chains are not observed above background. X-ray
signatures of electronic instabilities both in the chains and in
the layers must therefore be much weaker than the manifestations
of the short-range oxygen vacancy order we observe.

This conclusion is underscored by an intensity map of
{\mbox{YBa$_2$Cu$_4$O$_{8}$}, shown in Fig. \ref{fig:y124} on the
same scale as those of Fig. \ref{fig:ycado}. Since this material
does not sustain oxygen defects, diffuse scattering due to stripe
or charge density wave ordering should be much more easily visible
than in \ybco. However, no such intensity is observed at least
down to the intensity level of the diffuse peaks in the optimally
doped compound.

Finally, prompted by reports of large isotope effects on various
physical properties of high temperature superconductors
\cite{Ber03}, we have also investigated an optimally doped \ybcoo\
crystal in which $\rm ^{16}O$ was completely exchanged by $\rm
^{18}O$. The diffuse scattering pattern (not shown) was found to
be virtually identical to that of the $\rm ^{16}O$-rich material.
This excludes variations in oxygen short-range order as the origin
of the observed isotope effects, at least for the \ybco\ system.

In conclusion, we have observed a superstructure with a
four-unit-cell periodicity in $\rm (Y,Ca)Ba_2 Cu_3 O_{6+x}$
materials with $x \sim 0.95$. The superstructure involves atomic
displacements throughout the unit cell, but its origin can
unambiguously be attributed to a short-range ordering of oxygen
vacancies. This is supported by three independent observations:
the similarity of the diffuse scattering patterns of the \ybcoo\
and \ycado\ compounds, which have similar oxygen content but
different charge carrier concentrations; the absence of diffuse
reflections in the scattering pattern of the \yezv\ compound,
which contains no oxygen vacancies; and the persistence of the
diffuse intensities up to temperatures well above room
temperature.

Charge density modulations along the CuO chains of \ybco\ have been
addressed both in the experimental
\cite{Kra99,Gre00,Maki02,Derro02} and in the theoretical
\cite{Morr01} literature. Experimental evidence includes
variations of the NQR frequencies of both chain and plane
$^{63}$Cu nuclei, as well as spatially periodic modulations of the
tunnelling conductance in STM studies. These observations have
been interpreted in terms of either charge density wave
correlations \cite{Kra99,Gre00,Maki02} or Friedel-type
oscillations around oxygen defects in the chains
\cite{Derro02,Morr01}. Subtle modulations of the valence electron
density along the chains may be too weak to be observed by x-rays
in the presence of much stronger diffuse scattering from
short-range ordered oxygen defects (which involves all core
electrons). However, our observation of significant interchain
correlations in the positions of oxygen defects may lead to a more
quantitative understanding of the NQR and STM results. Indeed,
correlations between charge density modulations on different
chains observed in STM images \cite{Derro02} may well originate in
the correlations between oxygen vacancies reported here.

As an integral part of the lattice structure of optimally doped
$\rm YBa_2 Cu_3 O_{6+x}$, probably the most extensively
investigated high temperature superconductor -- stoichiometric
$\rm YBa_2 Cu_3 O_{7}$ is overdoped and difficult to prepare --
the oxygen ordering-induced superstructure may also be relevant
for the interpretation of a variety of other spectral features. In
particular, this component of the real lattice structure should be
taken into account when interpreting phonon anomalies in
underdoped \cite{Mook98} and optimally doped \cite{Pin03} $\rm
YBa_2 Cu_3 O_{6+x}$, which are expected to be sensitive to the
local lattice displacements reported here. Finally, based on our
findings we suggest to investigate short-range correlations
between oxygen defects and their potential impact on spectroscopic
features in other families of high temperature superconductors as
well.

We thank B. Dabrowski (NIU) for providing the \yezv\ sample and
J.C. Davis, S.C. Moss, and P. Wochner for valuable discussions.
Use of the Advanced Photon Source is supported by the U.S.
Department of Energy, Office of Science, Office of Basic Energy
Sciences, under contract W-31-109-Eng-38.

\end{document}